\def\today{
  \number\day \space \ifcase\month\or
  January\or February\or March\or April\or May\or June\or
  July\or August\or September\or October\or November\or December\fi
\space\number\year}
%
%
%
\def\unredoffs{} \def\redoffs{\voffset=-.31truein\hoffset=-.59truein}
\def\speclscape{\special{ps: landscape}}
%
%
%
%
\newbox\leftpage \newdimen\fullhsize \newdimen\hstitle \newdimen\hsbody
\tolerance=1000\hfuzz=2pt
\catcode`\@=11 
\def\bigans{b }
\def\answ{b }
%
\ifx\answ\bigans\message{(This will come out unreduced.}
\magnification=1200\unredoffs\baselineskip=16pt plus 2pt minus 1pt
\hsbody=\hsize \hstitle=\hsize 
\else\message{(This will be reduced.} \let\l@r=L
\magnification=1000\baselineskip=16pt plus 2pt minus 1pt \vsize=7truein
\redoffs \hstitle=8truein\hsbody=4.75truein\fullhsize=10truein\hsize=\hsbody
\output={\ifnum\pageno=0 
  \shipout\vbox{\speclscape{\hsize\fullhsize\makeheadline}
    \hbox to \fullhsize{\hfill\pagebody\hfill}}\advancepageno
  \else
  \almostshipout{\leftline{\vbox{\pagebody\makefootline}}}\advancepageno
  \fi}
\def\almostshipout#1{\if L\l@r \count1=1 \message{[\the\count0.\the\count1]}
      \global\setbox\leftpage=#1 \global\let\l@r=R
 \else \count1=2
  \shipout\vbox{\speclscape{\hsize\fullhsize\makeheadline}
      \hbox to\fullhsize{\box\leftpage\hfil#1}}  \global\let\l@r=L\fi}
\fi
%
\newcount\yearltd\yearltd=\year\advance\yearltd by -1900

\def\Title#1#2{\nopagenumbers\abstractfont\hsize=\hstitle\rightline{#1}%
\vskip 1in\centerline{\titlefont #2}\abstractfont\vskip .5in\pageno=0}
\def\Date#1{\vfill\leftline{#1}\tenpoint\supereject\global\hsize=\hsbody%
\footline={\hss\tenrm\folio\hss}}
%

\def\draftmode{\message{ DRAFTMODE }\def\draftdate{{\rm preliminary draft:
\number\month/\number\day/\number\yearltd\ \ \hourmin}}%
\headline={\hfil\draftdate}\writelabels\baselineskip=20pt plus 2pt minus 2pt
 {\count255=\time\divide\count255 by 60 \xdef\hourmin{\number\count255}
  \multiply\count255 by-60\advance\count255 by\time
  \xdef\hourmin{\hourmin:\ifnum\count255<10 0\fi\the\count255}}}
\def\nolabels{\def\wrlabeL##1{}\def\eqlabeL##1{}\def\reflabeL##1{}}
\def\writelabels{\def\wrlabeL##1{\leavevmode\vadjust{\rlap{\smash%
{\line{{\escapechar=` \hfill\rlap{\sevenrm\hskip.03in\string##1}}}}}}}%
\def\eqlabeL##1{{\escapechar-1\rlap{\sevenrm\hskip.05in\string##1}}}%
\def\reflabeL##1{\noexpand\llap{\noexpand\sevenrm\string\string\string##1}}}
\nolabels
%
\global\newcount\secno \global\secno=0
\global\newcount\meqno \global\meqno=1
\def\newsec#1{\global\advance\secno by1\message{(\the\secno. #1)}
\global\subsecno=0\eqnres@t\noindent{\bf\the\secno. #1}
\writetoca{{\secsym} {#1}}\par\nobreak\medskip\nobreak}
\def\eqnres@t{\xdef\secsym{\the\secno.}\global\meqno=1\bigbreak\bigskip}
\def\sequentialequations{\def\eqnres@t{\bigbreak}}\xdef\secsym{}
\global\newcount\subsecno \global\subsecno=0
\def\subsec#1{\global\advance\subsecno by1\message{(\secsym\the\subsecno. #1)}
\ifnum\lastpenalty>9000\else\bigbreak\fi
\noindent{\it\secsym\the\subsecno. #1}\writetoca{\string\quad
{\secsym\the\subsecno.} {#1}}\par\nobreak\medskip\nobreak}
\def\appendix#1#2{\global\meqno=1\global\subsecno=0\xdef\secsym{\hbox{#1.}}
\bigbreak\bigskip\noindent{\bf Appendix #1. #2}\message{(#1. #2)}
\writetoca{Appendix {#1.} {#2}}\par\nobreak\medskip\nobreak}
%
%
\def\eqnn#1{\xdef #1{(\secsym\the\meqno)}\writedef{#1\leftbracket#1}%
\global\advance\meqno by1\wrlabeL#1}
\def\eqna#1{\xdef #1##1{\hbox{$(\secsym\the\meqno##1)$}}
\writedef{#1\numbersign1\leftbracket#1{\numbersign1}}%
\global\advance\meqno by1\wrlabeL{#1$\{\}$}}
\def\eqn#1#2{\xdef #1{(\secsym\the\meqno)}\writedef{#1\leftbracket#1}%
\global\advance\meqno by1$$#2\eqno#1\eqlabeL#1$$}
%
\newskip\footskip\footskip14pt plus 1pt minus 1pt 
\def\footnotefont{\ninepoint}\def\f@t#1{\footnotefont #1\@foot}
\def\f@@t{\baselineskip\footskip\bgroup\footnotefont\aftergroup\@foot\let\next}
\setbox\strutbox=\hbox{\vrule height9.5pt depth4.5pt width0pt}
\global\newcount\ftno \global\ftno=0
\def\foot{\global\advance\ftno by1\footnote{$^{\the\ftno}$}}
%
\newwrite\ftfile
\def\footend{\def\foot{\global\advance\ftno by1\chardef\wfile=\ftfile
$^{\the\ftno}$\ifnum\ftno=1\immediate\openout\ftfile=foots.tmp\fi%
\immediate\write\ftfile{\noexpand\smallskip%
\noexpand\item{f\the\ftno:\ }\pctsign}\findarg}%
\def\footatend{\vfill\eject\immediate\closeout\ftfile{\parindent=20pt
\centerline{\bf Footnotes}\nobreak\bigskip\input foots.tmp }}}
\def\footatend{}
%
%
\global\newcount\refno \global\refno=1
\newwrite\rfile
\def\ref{[\the\refno]\nref}
\def\nref#1{\xdef#1{[\the\refno]}\writedef{#1\leftbracket#1}%
\ifnum\refno=1\immediate\openout\rfile=refs.tmp\fi
\global\advance\refno by1\chardef\wfile=\rfile\immediate
\write\rfile{\noexpand\item{#1\ }\reflabeL{#1\hskip.31in}\pctsign}\findarg}
\def\findarg#1#{\begingroup\obeylines\newlinechar=`\^^M\pass@rg}
{\obeylines\gdef\pass@rg#1{\writ@line\relax #1^^M\hbox{}^^M}%
\gdef\writ@line#1^^M{\expandafter\toks0\expandafter{\striprel@x #1}%
\edef\next{\the\toks0}\ifx\next\em@rk\let\next=\endgroup\else\ifx\next\empty%
\else\immediate\write\wfile{\the\toks0}\fi\let\next=\writ@line\fi\next\relax}}
\def\striprel@x#1{} \def\em@rk{\hbox{}}
\def\lref{\begingroup\obeylines\lr@f}
\def\lr@f#1#2{\gdef#1{\ref#1{#2}}\endgroup\unskip}

\def\addref#1{\immediate\write\rfile{\noexpand\item{}#1}} 
\def\startrefs#1{\immediate\openout\rfile=refs.tmp\refno=#1}
\def\xref{\expandafter\xr@f}\def\xr@f[#1]{#1}
\def\refs#1{\count255=1[\r@fs #1{\hbox{}}]}
\def\r@fs#1{\ifx\und@fined#1\message{reflabel \string#1 is undefined.}%
\nref#1{need to supply reference \string#1.}\fi%
\vphantom{\hphantom{#1}}\edef\next{#1}\ifx\next\em@rk\def\next{}%
\else\ifx\next#1\ifodd\count255\relax\xref#1\count255=0\fi%
\else#1\count255=1\fi\let\next=\r@fs\fi\next}
%

%
\newwrite\ffile\global\newcount\figno \global\figno=1
\def\fig{fig.~\the\figno\nfig}
\def\nfig#1{\xdef#1{fig.~\the\figno}%
\writedef{#1\leftbracket fig.\noexpand~\the\figno}%
\ifnum\figno=1\immediate\openout\ffile=figs.tmp\fi\chardef\wfile=\ffile%
\immediate\write\ffile{\noexpand\medskip\noexpand\item{Fig.\ \the\figno. }
\reflabeL{#1\hskip.55in}\pctsign}\global\advance\figno by1\findarg}
\def\xfig{\expandafter\xf@g}\def\xf@g fig.\penalty\@M\ {}
\def\figs#1{figs.~\f@gs #1{\hbox{}}}
\def\f@gs#1{\edef\next{#1}\ifx\next\em@rk\def\next{}\else
\ifx\next#1\xfig #1\else#1\fi\let\next=\f@gs\fi\next}
\newwrite\lfile
{\escapechar-1\xdef\pctsign{\string\%}\xdef\leftbracket{\string\{}
\xdef\rightbracket{\string\}}\xdef\numbersign{\string\#}}

\def\writestop{\def\writestoppt{\immediate\write\lfile{\string\pageno%
\the\pageno\string\startrefs\leftbracket\the\refno\rightbracket%
\string\def\string\secsym\leftbracket\secsym\rightbracket%
\string\secno\the\secno\string\meqno\the\meqno}\immediate\closeout\lfile}}
\def\writestoppt{}\def\writedef#1{}
\def\seclab#1{\xdef #1{\the\secno}\writedef{#1\leftbracket#1}\wrlabeL{#1=#1}}
\def\subseclab#1{\xdef #1{\secsym\the\subsecno}%
\writedef{#1\leftbracket#1}\wrlabeL{#1=#1}}
\newwrite\tfile \def\writetoca#1{}
\def\leaderfill{\leaders\hbox to 1em{\hss.\hss}\hfill}
\def\writetoc{\immediate\openout\tfile=toc.tmp
   \def\writetoca##1{{\edef\next{\write\tfile{\noindent ##1
   \string\leaderfill {\noexpand\number\pageno} \par}}\next}}}

%
\catcode`\@=12 
%
\edef\tfontsize{\ifx\answ\bigans scaled\magstep3\else scaled\magstep4\fi}
\font\titlerm=cmr10 \tfontsize \font\titlerms=cmr7 \tfontsize
\font\titlermss=cmr5 \tfontsize \font\titlei=cmmi10 \tfontsize
\font\titleis=cmmi7 \tfontsize \font\titleiss=cmmi5 \tfontsize
\font\titlesy=cmsy10 \tfontsize \font\titlesys=cmsy7 \tfontsize
\font\titlesyss=cmsy5 \tfontsize \font\titleit=cmti10 \tfontsize
\skewchar\titlei='177 \skewchar\titleis='177 \skewchar\titleiss='177
\skewchar\titlesy='60 \skewchar\titlesys='60 \skewchar\titlesyss='60
\def\titlefont{\def\rm{\fam0\titlerm}
\textfont0=\titlerm \scriptfont0=\titlerms \scriptscriptfont0=\titlermss
\textfont1=\titlei \scriptfont1=\titleis \scriptscriptfont1=\titleiss
\textfont2=\titlesy \scriptfont2=\titlesys \scriptscriptfont2=\titlesyss
\textfont\itfam=\titleit \def\it{\fam\itfam\titleit}\rm}
 \ifx\answ\bigans\else scaled\magstep1\fi
\ifx\answ\bigans\def\abstractfont{\tenpoint}\else
\font\abssl=cmsl10 scaled \magstep1
\font\absrm=cmr10 scaled\magstep1 \font\absrms=cmr7 scaled\magstep1
\font\absrmss=cmr5 scaled\magstep1 \font\absi=cmmi10 scaled\magstep1
\font\absis=cmmi7 scaled\magstep1 \font\absiss=cmmi5 scaled\magstep1
\font\abssy=cmsy10 scaled\magstep1 \font\abssys=cmsy7 scaled\magstep1
\font\abssyss=cmsy5 scaled\magstep1 \font\absbf=cmbx10 scaled\magstep1
\skewchar\absi='177 \skewchar\absis='177 \skewchar\absiss='177
\skewchar\abssy='60 \skewchar\abssys='60 \skewchar\abssyss='60
\def\abstractfont{\def\rm{\fam0\absrm}
\textfont0=\absrm \scriptfont0=\absrms \scriptscriptfont0=\absrmss
\textfont1=\absi \scriptfont1=\absis \scriptscriptfont1=\absiss
\textfont2=\abssy \scriptfont2=\abssys \scriptscriptfont2=\abssyss
\textfont\itfam=\bigit \def\it{\fam\itfam\bigit}\def\footnotefont{\tenpoint}%
\textfont\slfam=\abssl \def\sl{\fam\slfam\abssl}%
\textfont\bffam=\absbf \def\bf{\fam\bffam\absbf}\rm}\fi
\def\tenpoint{\def\rm{\fam0\tenrm}
\textfont0=\tenrm \scriptfont0=\sevenrm \scriptscriptfont0=\fiverm
\textfont1=\teni  \scriptfont1=\seveni  \scriptscriptfont1=\fivei
\textfont2=\tensy \scriptfont2=\sevensy \scriptscriptfont2=\fivesy
\textfont\itfam=\tenit \def\it{\fam\itfam\tenit}\def\footnotefont{\ninepoint}%
\textfont\bffam=\tenbf \def\bf{\fam\bffam\tenbf}\def\sl{\fam\slfam\tensl}\rm}
\font\ninerm=cmr9 \font\sixrm=cmr6 \font\ninei=cmmi9 \font\sixi=cmmi6
\font\ninesy=cmsy9 \font\sixsy=cmsy6 \font\ninebf=cmbx9
\font\nineit=cmti9 \font\ninesl=cmsl9 \skewchar\ninei='177
\skewchar\sixi='177 \skewchar\ninesy='60 \skewchar\sixsy='60
\def\ninepoint{\def\rm{\fam0\ninerm}
\textfont0=\ninerm \scriptfont0=\sixrm \scriptscriptfont0=\fiverm
\textfont1=\ninei \scriptfont1=\sixi \scriptscriptfont1=\fivei
\textfont2=\ninesy \scriptfont2=\sixsy \scriptscriptfont2=\fivesy
\textfont\itfam=\ninei \def\it{\fam\itfam\nineit}\def\sl{\fam\slfam\ninesl}%
\textfont\bffam=\ninebf \def\bf{\fam\bffam\ninebf}\rm}
%
%

\hyphenation{anom-aly anom-alies coun-ter-term coun-ter-terms}
\def\inv{^{\raise.15ex\hbox{${\scriptscriptstyle -}$}\kern-.05em 1}}

\def\Dsl{\,\raise.15ex\hbox{/}\mkern-13.5mu D} 
\def\dsl{\raise.15ex\hbox{/}\kern-.57em\partial}

\font\bigit=cmti10 scaled \magstep1
\def\lspace{\ifx\answ\bigans{}\else\qquad\fi}
\def\lbspace{\ifx\answ\bigans{}\else\hskip-.2in\fi} 
\def\boxeqn#1{\vcenter{\vbox{\hrule\hbox{\vrule\kern3pt\vbox{\kern3pt
	\hbox{${\displaystyle #1}$}\kern3pt}\kern3pt\vrule}\hrule}}}
\def\mbox#1#2{\vcenter{\hrule \hbox{\vrule height#2in
		\kern#1in \vrule} \hrule}}  
%
   
\def\CL{{\cal L}}

\def\grad#1{\,\nabla\!_{{#1}}\,}

\def\darr#1{\raise1.5ex\hbox{$\leftrightarrow$}\mkern-16.5mu #1}

\def\roughly#1{\raise.3ex\hbox{$#1$\kern-.75em\lower1ex\hbox{$\sim$}}}


\hyphenation{anom-aly anom-alies coun-ter-term coun-ter-terms
dif-feo-mor-phism dif-fer-en-tial super-dif-fer-en-tial dif-fer-en-tials
super-dif-fer-en-tials reparam-etrize param-etrize reparam-etriza-tion}


%
%
%
\newwrite\tocfile\global\newcount\tocno\global\tocno=1
\ifx\bigans\answ \def\tocline#1{\hbox to 320pt{\hbox to 45pt{}#1}}
\else\def\tocline#1{\line{#1}}\fi
\def\toclead{\leaders\hbox to 1em{\hss.\hss}\hfill}
\def\tnewsec#1#2{\xdef #1{\the\secno}\newsec{#2}
\ifnum\tocno=1\immediate\openout\tocfile=toc.tmp\fi\global\advance\tocno
by1
{\let\the=0\edef\next{\write\tocfile{\medskip\tocline{\secsym\ #2\toclead\the
\count0}\smallskip}}\next}
}
\def\tnewsubsec#1#2{\xdef #1{\the\secno.\the\subsecno}\subsec{#2}
\ifnum\tocno=1\immediate\openout\tocfile=toc.tmp\fi\global\advance\tocno
by1
{\let\the=0\edef\next{\write\tocfile{\tocline{ \ \secsym\the\subsecno\
#2\toclead\the\count0}}}\next}
}
\def\tappendix#1#2#3{\xdef #1{#2.}\appendix{#2}{#3}
\ifnum\tocno=1\immediate\openout\tocfile=toc.tmp\fi\global\advance\tocno
by1
{\let\the=0\edef\next{\write\tocfile{\tocline{ \ #2.
#3\toclead\the\count0}}}\next}
}
%
%

%
%
%
%

%
\long\def\optional#1{}
%
%
\let\narrowequiv=\equiv
\def\equiv{\;\narrowequiv\;}

\def\tilde{\widetilde}
\fontdimen16\tensy=2.7pt\fontdimen17\tensy=2.7pt 



%

%
%

\def\CL{{\cal L}}

%
%
%
\def\boxit#1#2{
        $$\vcenter{\vbox{\hrule\hbox{\vrule\kern3pt\vbox{\kern3pt
	\hbox to #1truein{\hsize=#1truein\vbox{#2}}\kern3pt}\kern3pt\vrule}
        \hrule}}$$
}


%




\def\splitexact#1#2{\mathrel{\mathop{\null{
\lower4pt\hbox{$\rightarrow$}\atop\raise4pt\hbox{$\leftarrow$}}}\limits
^{#1}_{#2}}}

%
%

%
%
%
%

\def\dd{\mskip 1.3mu{\rm d}\mskip .7mu} 



%
%

\def\IM{isomorphism}

%
%

\font\blackboard=msym10 \font\blackboards=msym7   
\font\blackboardss=msym5
\newfam\black
\textfont\black=\blackboard
\scriptfont\black=\blackboards
\scriptscriptfont\black=\blackboardss

\font\gothic=eufm10 \font\gothics=eufm7
\font\gothicss=eufm5
\newfam\gothi
\textfont\gothi=\gothic
\scriptfont\gothi=\gothics
\scriptscriptfont\gothi=\gothicss

\font\curlyfont=eusm10 \font\curlyfonts=eusm7
\font\curlyfontss=eusm5
\newfam\curly
\textfont\curly=\curlyfont
\scriptfont\curly=\curlyfonts
\scriptscriptfont\curly=\curlyfontss


\global\newcount\figno \global\figno=1
\newwrite\ffile
\def\pfig#1#2{Fig.~\the\figno\pnfig#1{#2}}
\def\pnfig#1#2{\xdef#1{Fig. \the\figno}%
\ifnum\figno=1\immediate\openout\ffile=figs.tmp\fi%
\immediate\write\ffile{\noexpand\item{\noexpand#1\ }#2}%
\global\advance\figno by1}

%
%
\def\tfig#1{Fig.~\the\figno\xdef#1{Fig.~\the\figno}\global\advance\figno by1}








%
%

%


\def\inbar{\,\vrule height1.5ex width.4pt depth0pt}
\def\IB{\relax{\rm I\kern-.18em B}}
\def\IC{\relax\hbox{$\inbar\kern-.3em{\rm C}$}}
\def\ID{\relax{\rm I\kern-.18em D}}
\def\IE{\relax{\rm I\kern-.18em E}}
\def\IF{\relax{\rm I\kern-.18em F}}
\def\IG{\relax\hbox{$\inbar\kern-.3em{\rm G}$}}
\def\IH{\relax{\rm I\kern-.18em H}}
\def\II{\relax{\rm I\kern-.18em I}}
\def\IK{\relax{\rm I\kern-.18em K}}
\def\IL{\relax{\rm I\kern-.18em L}}
\def\IM{\relax{\rm I\kern-.18em M}}
\def\IN{\relax{\rm I\kern-.18em N}}
\def\IO{\relax\hbox{$\inbar\kern-.3em{\rm O}$}}
\def\IP{\relax{\rm I\kern-.18em P}}
\def\IQ{\relax\hbox{$\inbar\kern-.3em{\rm Q}$}}
\def\IR{\relax{\rm I\kern-.18em R}}
\font\cmss=cmss10 \font\cmsss=cmss10 at 10truept
\def\IZ{\relax\ifmmode\mathchoice
{\hbox{\cmss Z\kern-.4em Z}}{\hbox{\cmss Z\kern-.4em Z}}
{\lower.9pt\hbox{\cmsss Z\kern-.36em Z}}
{\lower1.2pt\hbox{\cmsss Z\kern-.36em Z}}\else{\cmss Z\kern-.4em Z}\fi}
\def\IGa{\relax\hbox{${\rm I}\kern-.18em\Gamma$}}
\def\IPi{\relax\hbox{${\rm I}\kern-.18em\Pi$}}
\def\ITh{\relax\hbox{$\inbar\kern-.3em\Theta$}}
\def\IOm{\relax\hbox{$\inbar\kern-3.00pt\Omega$}}

\def\dot{\!\cdot\!}
\def\cross{\!\times\!}

\def\kbT{k_{\scriptscriptstyle\rm B}T}

\def\dnb{\delta{\bf n}}

\def\bold#1{\setbox0=\hbox{$#1$}%
     \kern-.010em\copy0\kern-\wd0
     \kern.025em\copy0\kern-\wd0
     \kern-.020em\raise.0200em\box0 }

\def\grad{\bold{\nabla}_{\!\!\scriptscriptstyle \perp}}

\nfig\fonn{Screw dislocation in a capillary geometry.  The dark center
line is the screw dislocation.}
\nfig\fone{Reduced critical field $\overline{H}$ vs. reduced capillary radius
$\overline{R}$ for $\overline{W}=.30,.54,.96,1.7,3.0.$}

\lref\ZU{S.~Kralj and S.~\v Zumer, Phys. Rev. E. {\bf 51} (1995) 366.}
\lref\SLZ{A.~Sugimura, G.R.~Luckhurst and O.-Y.~Zhong-can, Phys. Rev. E {\bf
52}
(1995) 681.}
\lref\ZUII{M.~Ambro\v zi\v c and S.~\v Zumer, Phys. Rev. E {\bf 54} (1996)
5187.}
\lref\DGP{P.G.~de Gennes and J.~Prost, {\sl The Physics of Liquid Crystals},
Second Edition, Chap. VII (Oxford University Press, New York, 1993).}
\lref\DGS{P.G.~de Gennes, Solid State Commun. {\bf 14}, 997 (1973).}
\lref\CL{See, for instance, P.M.~Chaikin and T.C.~Lubensky, {\sl Principles of
Condensed Matter Physics},
(Cambridge University Press, Cambridge, 1995).}
\lref\RL{S.R.~Renn and T.C.~Lubensky, Phys. Rev. A {\bf 38} (1988) 2132;
{\bf 41} (1990) 4392.}
\lref\DIS{J.~Goodby, M.A.~Waugh, S.M.~Stein. R.~Pindak, and J.S.~Patel,
Nature {\bf 337} (1988) 449; J. Am. Chem. Soc. {\bf 111} (1989) 8119;
G.~Strajer, R.~Pindak, M.A.~Waugh, J.W.~Goodby, and J.S.~Patel, Phys. Rev.
Lett.
{\bf 64} (1990) 13; K.J.~Ihn, J.A.N.~Zasadzinski, R.~Pindak, A.J.~Slaney, and
J.
{}~Goodby,
Science {\bf 258} (1992) 275.}
\lref\NGN{L.~ Navailles, C.W.~Garland and H.T.~Nguyen,
J. Phys. II France {\bf 6} (1996) 1243.}
\lref\buckle{For the case of {\sl hexatic} membranes, this was studied in
S.~Seung and D.R.~Nelson,
Phys. Rev. A {\bf 38} (1989) 1005; S.~Seung, Ph.D. Thesis, Harvard
University (1990).}

\Title{}{\vbox{\centerline{Determining the Anchoring Strength of a}\vskip 2pt
\centerline{Capillary Using Topological Defects}}}

\centerline{
Randall D. Kamien\footnote{$^\dagger$}{email:
{\tt kamien@lubensky.physics.upenn.edu}}}
\centerline{\sl Department of Physics and Astronomy, University of
Pennsylvania,
Philadelphia, PA 19104}
\smallskip\centerline{and}\smallskip
\centerline{Thomas R. Powers\footnote{$^\ddagger$}{email: {\tt
powers@davinci.physics.arizona.edu}}}
\centerline{\sl Department of Physics, University of Arizona, Tucson, AZ 85721}
\vskip .3in
We consider a smectic-$A^*$ in a capillary with surface anchoring
that favors parallel alignment.  If the bulk phase of the
smectic is the standard twist-grain-boundary phase of chiral
smectics, then there will be a critical radius below which
the smectic will not have any topological defects.  Above this
radius a single screw dislocation in the center of the capillary will
be favored.  Along with surface anchoring, a magnetic field will
also suppress the formation of a screw dislocation.
In this note,
we calculate the critical field at which a defect is energetically preferred
as a function of the surface anchoring strength and the capillary radius.
Experiments at a few different radii could thus determine the anchoring
strength.
\Date{17 December 1996; Revised 1 February 1997}

Boundary conditions play an essential role in liquid crystal physics and
they cannot be taken with a cavalier attitude.
Even if the boundary is very far away, surface effects in
liquid crystals can be very important because
of typically long-range, algebraic correlations in these soft materials.
Indeed, in device applications one is often
interested in the surface effects on bulk ordering: such effects are
a key element in the twisted nematic display \CL .

In this note we
consider a smectic liquid crystal which, in bulk, would form a
Renn-Lubensky \RL\ twist-grain-boundary
(TGB) phase \DIS .   We will show that if this smectic is confined to the
classic
capillary geometry \refs{\ZU,\ZUII}\ with the layer normals parallel to the
capillary axis then
for sufficiently small radii an undefected smectic-$A$ phase
persists while for larger radii screw dislocations can enter as shown in
figure 1.
While this, in principle, can give a very clean determination of the
anchoring strength $W$ it is rather impractical to do an experiment
on a sequence of capillaries with small differences in their radial size.
Instead, we show how the imposition of a magnetic field also suppresses
defect formation.  In this case, at fixed radii, the applied field may be
scanned and the critical field may be determined.  Doing this measurement
at a few radii should make it possible to determine $W$.

We start with the free energy of a smectic confined to a region $\Omega$ in
the $xy$-plane, and assume that the capillary is infinite in the $z$ direction.
We further assume there is an infinite energy cost for the director to have
a radial component at the surface.
The appropriate free energy to quadratic order is then \DGS\
\eqn\ei{\eqalign{F&=F_{\rm bulk}+F_{\rm surface}\cr
&=\int \dd z\int_\Omega \dd^2\!x\,\bigg\{{B\over 2}\left(\grad u +\dnb\right)^2
+ {B_z\over 2}
(\partial_z u)^2
+{K_1\over 2}\left(\grad\dot\dnb\right)^2 + {K_2\over
2}\left(\grad\cross\dnb\right)^2\cr
&\qquad\qquad\quad\qquad\qquad+{K_3\over 2}\left(\partial_z\dnb\right)^2
- {\chi\over 2}\left({\bf H}\dot{\bf
n}\right)^2\bigg\}\cr
&-\int \dd z\bigg\{\int_{\partial\Omega} \dd\ell \bigg[K_2q_0\dnb\dot{\bf T}+
{K_{24}\over 2}{\bf N}
\dot\left[{\bf n}\left(\bold{\nabla}\dot{\bf n}\right) - \left({\bf n}\dot
\bold{\nabla}\right){\bf n}\right]\bigg] -\int_{\partial_+\Omega}
\dd\ell\bigg[{W\over 2}\left({\bf n}_\phi\right)^2\bigg]\bigg\}.\cr}}
$\partial \Omega$ denotes the total boundary of the smectic region,
including any boundaries at defects, while $\partial_+\Omega$ includes
only the boundary with the capillary;
${\bf N}$ is the (outward pointing) normal to the surface and
${\bf T}$ is the surface tangent perpendicular to $\hat z$.
The smectic order parameter is $\psi=\vert\psi\vert e^{i2\pi(z+u)/a}$,
and thus
the phase of the mass-density wave is $u$. The molecular director
is ${\bf n}=\hat z
\sqrt{1-\delta n^2}+\dnb$ with $\hat z \dot \dnb=0$.
The $K_i$ are the
Frank elastic constants, $B$ and $B_z$ are proportional to the smectic order
parameter squared $\vert\psi\vert^2$, $K_{24}$ is the saddle-splay elastic
constant \DGP , $q_0$
is the equilibrium cholesteric pitch, $\chi$ is
the negative diamagnetic susceptibility, ${\bf H}=H\hat z$ is a
magnetic field, and
$W$ is the anchoring strength which we want to determine.  In \ei\ we have
used the fact that the chiral bulk term
$K_2q_0\grad\cross\dnb$ can be rewritten as a surface term.
While we are not explicitly interested in the saddle-splay term, there is
no way to remove it from the problem---when there is a surface it must
be included \refs{\SLZ,\ZUII}.  It too serves to promote or hinder
the formation of a central screw dislocation.

We note by symmetry along $\hat z$ that
the problem can be reduced to a {\sl two-dimensional} problem and thus the
fields
will have no $z$ dependence.  The field configurations which minimize the bulk
free energy satisfy the Euler-Lagrange equations:
\eqna\eiii{$$\eqalignno{
0&=\nabla_{\!\!\scriptscriptstyle\perp}^2u+\grad\dot\dnb&\eiii a\cr
0&=B\left(\grad u +\dnb\right)
+ \chi H^2\dnb-K_1\grad\grad\dot\dnb
- K_2(\nabla^2_{\!\!\scriptscriptstyle\perp}\dnb-\grad\grad\dot\dnb)&\eiii b
\cr}$$}

\noindent The saddle-splay term can lead to a buckling instability toward a
configuration
with a non-zero divergence of $\dnb$.
In this note we will limit our attention to the effect of the anchoring
term on defect formation;  in our quadratic approximation the transverse
distortions due to the defect decouple from the longitudinal distortions
due to buckling.
We will therefore assume $\grad\dot\;\dnb=0$ .
It would be an interesting
extension of this work to study the nonlinear theory to understand
buckling and its interplay with defects ({\it cf.} \buckle).

We write this constraint in cylindrical co\"ordinates as
\eqn\eiv{{1\over \rho}\partial_{\rho}\left(\rho\delta n_\rho\right) +
{1\over\rho}\partial_\phi\delta n_\phi=0.}
Using rotational invariance, we can assume that the nematic field $\dnb$ is
independent of $\phi$.  In this case \eiv\ implies that $\rho\delta n_\rho$ is
constant.  Since $\delta n$ must be well defined everywhere, including
$\rho=0$, the constant
must be $0$.  Therefore $\dnb=\delta n_\phi(\rho)\hat\phi$.

To find the minimum energy solution we must take into account the boundary
terms.  The outer surface energy in \ei\ may be expanded
up to quadratic order in $\dnb$:
\eqn\eii{F_{\rm surface} = \int \dd
z\int_{\partial_{+}\Omega}\dd\ell\left[\delta n_\phi(R)\right]^2
\left[{W\over 2}-{K_{24}\over 2 R}\right]-K_2 q_0 \int
\dd z\int_{\partial_+\Omega}\dd\ell\,\delta n_\phi(R),} where $R$ is the
capillary radius.  We work in the type II \DGS\ limit in which the twist
penetration depth $\lambda_0=\sqrt{K_2/B}$ is much bigger than the coherence
length $\xi$; in this limit the contribution of the inner boundary terms
turns out to be subleading.
We may now calculate the minimum bulk energy for fixed $\delta
n_\phi(R)$ and get an effective free energy $F\left(\delta n_\phi(R)\right)$
which we may finally minimize over the number $\delta n_\phi(R)$. At this
point we identify the {\sl magnetic} twist penetration depth
$\lambda_H^2=K_2/(B+\chi H^2)$.  We can solve
\eiii{}\ by introducing ${\bf Q}=\dnb + (\lambda_H^2/\lambda_0^2)\grad u$ \CL\
so that
\eqn\ev{\nabla^2_{\!\!\scriptscriptstyle\perp}{\bf Q} - {1\over\lambda_H^2}{\bf
Q}=0.}
Note that \eiii{b}\ implies that $u$ is independent of $\rho$ and thus
${\bf Q}$ only has components in the $\hat\phi$ direction.  In cylindrical
co\"ordinates
\ev\ reduces to Bessel's modified equation with index $\nu=1$ and
so
\eqn\evi{Q_\phi(\rho) = C_1K_1(\rho/\lambda_H) + C_2I_1(\rho/\lambda_H).}
We determine $C_1$ by insisting that $\dnb$ be regular at the origin.  The
origin
is, in fact, the {\sl other boundary}.  In particular, we might consider
the possibility that there is a defect at the center of the capillary.
Recall that a screw dislocation is a layer configuration in which
$u=am\phi/(2\pi)$ where
$a$ is the layer spacing and $m$ is an integer.  Note that since $u$ is
independent of $\rho$, $u=am\phi/(2\pi)$ is the most general
solution of the
equations of equilibrium with $\grad\dot\;\dnb=0$.
This implies that $\grad u =
\hat\phi{am\over(2\pi\rho)}$ and is thus singular at the origin.  Since
$I_1(0)=0$,
we have, near $\rho=0$
\eqn\evii{\rho\delta n_\phi(0) + {\lambda_H^2\over\lambda_0^2}{am\over 2\pi} =
C_1\lambda_H,}
and so $C_1=am\lambda_H/(2\pi\lambda_0^2)$.  Thus the only free parameter in
the bulk free energy is $C_2$ which is determined by demanding
$\delta n_\phi(R)\equiv\delta n_0$:
\eqn\eviii{C_2=  {\delta n_0 + am\lambda_H/(2\pi\lambda_0^2)\left[(\lambda_H/R)
 - K_1(R/\lambda_H)\right]
\over I_1(R/\lambda_H)} .}
Taking the core size to be $\xi$ and working in the regime where
$R\gg\lambda_0,\lambda_H\gg\xi$ we
find that the bulk strain energy to leading order in $\lambda_H/R$ is
\eqn\eix{\eqalign{
F_{\rm bulk}(\delta n_0)/L &\approx m^2 E_{\rm core} +
\chi H^2 {a^2m^2\lambda_H^2\over 4\pi\lambda_0^2}\ln(R/\xi)
+K_2{a^2m^2\lambda_H^2 \over 4\pi\lambda_0^4}\ln{\lambda_H
\over \xi}\cr
&+\delta n_0^2K_2\pi{R\over\lambda_H}
+\delta n_0K_2{\lambda_Ham\over\lambda_0^2},\cr}}
where $m^2E_{\rm core}$ is the energy cost per unit
length of destroying the smectic order at
the defect core.
Note that the strain energy depends not only on the boundary value of
$\dnb = \delta n_0\hat\phi$, but also on the strength
of the defect $m$.  When minimizing, we must minimize over
{\sl both} variables.  Choosing
the minimum over $m$ will indicate whether or not there is a
screw dislocation at all.

Adding this energy to the surface energy \eii, we minimize
over $\delta n_0$ to find the minimum energy $\tilde F$.
To determine the relative magnitudes of the terms,
we estimate
$K_2\approx K_{24}={\cal O}({\kbT\over\xi})$ and $a\approx\xi$.
In NMR studies \ref\Doane{G.P.~Crawford, D.W.~Allender, J.W.~Doane,
M. Vilfan, and I. Vilfan, Phys. Rev. A {\bf 44} (1991) 2570. }\
it has been seen that $W/K_{24}\approx (28\ {\rm nm})^{-1}$
for non-chiral nematic liquid crystal in
small capillaries in Nuclepore; we therefore expect that
$K_{24}/R \ll W$  for radii in the range of $R\sim 10\ \mu{\rm m}$.
Thus we find a tipping of
\eqn\etip{\delta n_0\approx{K_2 q_0\over W}{1-{\lambda_H m a\over 2\pi
\lambda_0^2 q_0 R}
\over 1+ {K_2\over W\lambda_H}}.}
For large $R$ the tipping angle at the
capillary wall is {\sl independent} of $R$ and $m$.  This is because
the director can lower the chiral energy by tipping at the wall even in
the absence of a defect; the director will align with the layer normal at
a distance of a penetration depth from the wall.  Therefore we must subtract
off the strain energy of the zero defect state to arrive at the dislocation
energy
\eqn\exi{\eqalign{\tilde F/L &\approx m^2\left\{
E_{\rm core} +
\chi H^2{a^2\lambda_H^2\over 4\pi\lambda_0^2}\ln{R\over\xi} +
K_2{a^2\lambda_H^2\over 4\pi\lambda_0^4}\ln{\lambda_H\over\xi}
\right\}+ m{K_2^2aq_0\lambda_H\over
(W+{K_2\over\lambda_H})\lambda_0^2}
\cr}}
We have kept the subleading term $K_2/(W\lambda_H)$ so that our expressions
have sensible large $H$ behaviour,
{\it e.g.} $\delta n_0\to 0$, as $H\to\infty$,
even though in this limit we leave the type II regime.
Each term of \exi\ has a simple interpretation: the first three
terms are the energy of a screw dislocation in a bulk sample subject to a
magnetic field, and the last term is the usual chiral term of a bulk
defect reduced by a factor involving the anchoring.

Note that if $q_0=0$ then the minimum energy is at $m=0$, in other
words, {\sl no defect}.
It is easy to see that there are critical
values of $H^2$, $R$ and $q_0$ for which the free
energy will be minimized for $m\ne 0$.
First we consider zero magnetic field.
The sign of the Burgers vector is determined by the
sign of $q_0$; we assume that $q_0>0$.
The critical value of the chirality needed to get a defect
with $m=-1$ in a very large capillary ($R\to\infty$) is increased from its
bulk value by the anchoring:
\eqn\qcrit{q_{0{\rm c}}\approx q_{0{\rm c, bulk}}
\left({W\lambda_0\over K_2}+1\right),}
where
\eqn\eqcc{q_{0{\rm c, bulk}}={m\over a}\left[{E_{\rm core}\over K_2}
+{a^2 \over 4\pi\lambda_0^2}\log(\lambda_0/\xi)\right].}
In general the critical chirality depends on the radius.  Equivalently,
there is a critical
radius $R_{\rm c}$ depending on $q_0$ such that if $R<R_{\rm c}$, no
defect will occur.  Unless $q_0$ is
very close to its critical value, this radius is
typically the size of the penetration depth, in which case our expressions
are not valid.  We can, of course, compute $R_{\rm c}$ and its dependence
on $W$ by dropping the assumption $R\gg\lambda_0,$ but as noted above
$R_{\rm c}$ is not so easy to determine experimentally.

Now consider subjecting the sample to a magnetic field.  For a capillary
with radius $R>R_{\rm c}$, there is a critical value $H_{\rm c}(R,W)$
of the field above which there will be no defect; $H_{\rm c}(R,W)$ is the
number such that the right hand side of \exi\ is zero for $m\ne 0$.
While an analytic expression for
$H=H_c$ is complicated, we can
easily plot $H_c(R,W)$ for a system in which the elastic constants are known.
Thus a given liquid crystal system may be studied in
only a few capillaries in order to determine which value of $W$ fits the
observed formation of a defect.
In fact, due to the logarithmic dependence of the magnetic strain energy
on the radius, the critical field is a very weak function of $R$ so that
just one measurement may suffice to determine $W$.
In figure 2 we show $H_c$ vs. $R$ for
various $W$, with $R$ ranging over typical capillary sizes
of $10$--$100\ \mu{\rm
m}$; $\overline{R}\equiv R/\lambda_0$,
$\overline{H}^2\equiv\chi H^2/B$,
$\overline{W}\equiv W\lambda_0/K_2$, and $q_0=11.\ \mu{\rm m}^{-1}$.
$\overline{W}$ ranges from $0.3$ to $3.0.$  We have taken $\lambda_0=0.1
\ \mu{\rm m}$, $4\pi\lambda_0^2E_{\rm core}/(a^2 K_2)=1$, and
$\lambda_0/\xi=10$.  Note that the magnetic
fields in the ranges we consider do not spoil the type II behavior
since $\lambda_H>\xi$.

In summary we have proposed a set of experiments that could be
performed on liquid crystals that form TGB phases, such as
{\sl 3-fluoro-4(1-methylheptyloxy)4$'$
-(4{$'$}{$'$}-alkoxy-2{$'$}{$'$},3{$'$}{$'$}-difluorobenzyoyloxy)tolane}
($nF_2BTFO_1M_7$)
\NGN\ which could, in principle, measure the anchoring strength of the
capillary wall.  Throughout we have neglected the dynamics of the formation
of screw dislocations from an undefected sample.  The nucleation problem
could be avoided experimentally through annealing.  It would be an interesting
extension of this work, however, to understand defect formation.  This
task would be considerably easier and tremendously more reliable
with some experimental input.

It is a pleasure to acknowledge stimulating discussions with
T.C.~Lubensky, P.L.~Taylor and S.~\v Zumer.
The authors acknowledge the hospitality of the Aspen Center for Theoretical
Physics, where some of this work was done.
RDK was supported by NSF Grants DMR94-23114 and DMR91-22645,
and TRP by NSF Grant DMR93-50227.

\footatend\vfill\supereject\immediate\closeout\rfile\writestoppt
\baselineskip=14pt\centerline{{\bf References}}\bigskip{\frenchspacing%
\parindent=20pt\escapechar=` \input refs.tmp\vfill\eject}\nonfrenchspacing
\vfill\eject\immediate\closeout\ffile
\centerline{{\bf Figure Captions}}\bigskip\frenchspacing%
\input figs.tmp\vfill\eject\nonfrenchspacing

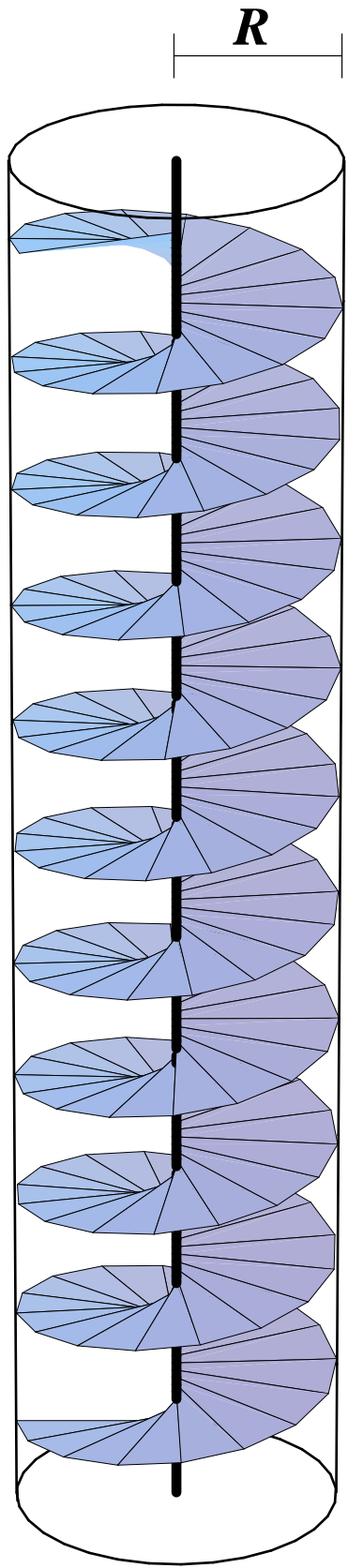
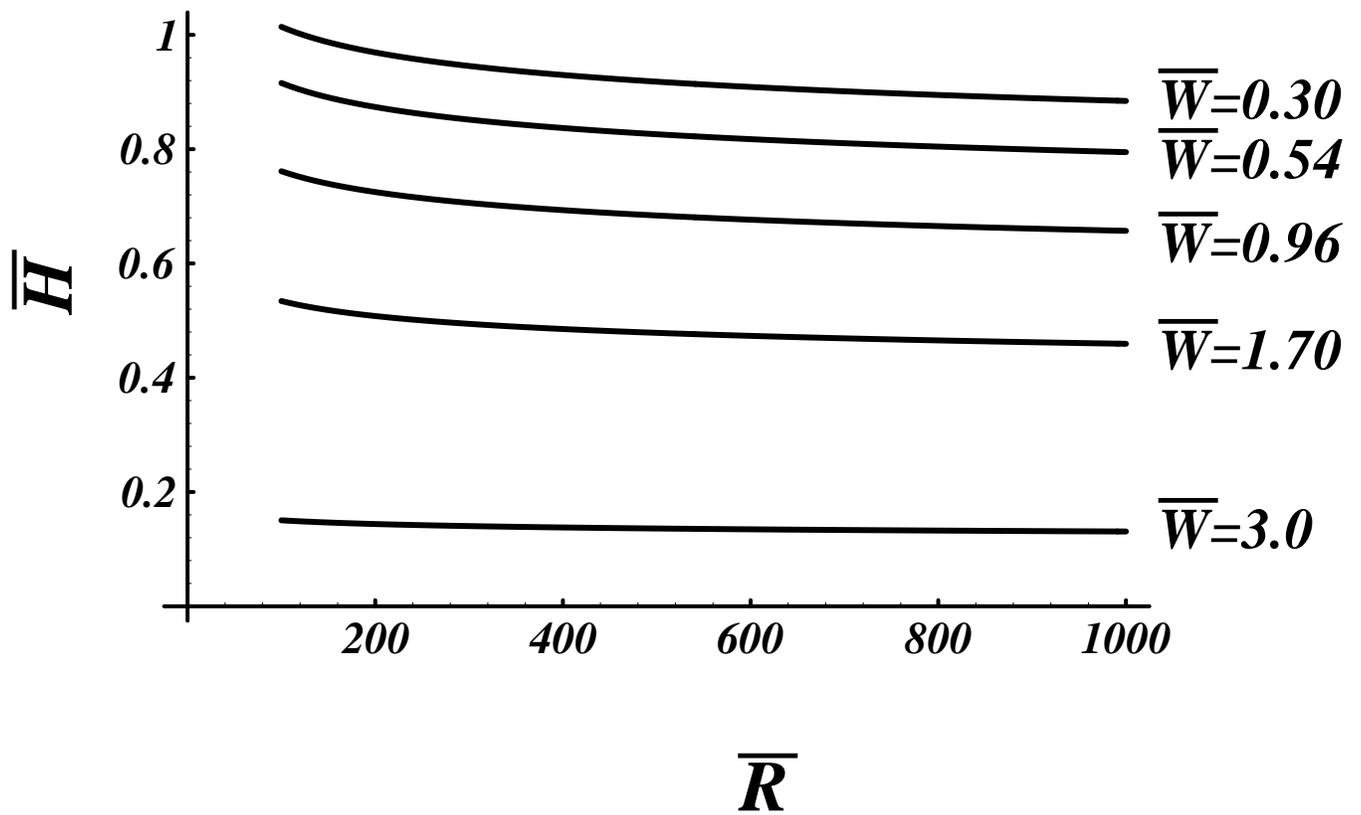
\bye